\documentclass[%
reprint,
superscriptaddress,
showpacs,
amsmath,amssymb,
prl,
floatfix, ]%
{revtex4-1}

\usepackage{color}
\usepackage{CJK}

\usepackage{graphicx}% Include figure files
\usepackage{dcolumn}% Align table columns on decimal point
\usepackage{bm}% bold math
\usepackage[dvipdfmx,bookmarks=true,colorlinks,%
            citecolor=blue,linkcolor=blue,anchorcolor=blue,filecolor=blue,urlcolor=blue,%
           ]{hyperref}

\begin{document}

%\begin{CJK*}{GBK}{}

\title{Pseudospin symmetry in single particle resonant states}

\author{Bing-Nan Lu}%(ÂÀ±þéª)}%
%\email{bnlu@itp.ac.cn}
 \affiliation{State Key Laboratory of Theoretical Physics,
              Institute of Theoretical Physics, Chinese Academy of Sciences, Beijing 100190, China}
\author{En-Guang Zhao}%(ÕÔ¶÷¹ã)}%
%\email{egzhao@mail.itp.ac.cn}
 \affiliation{State Key Laboratory of Theoretical Physics,
              Institute of Theoretical Physics, Chinese Academy of Sciences, Beijing 100190, China}
 \affiliation{Center of Theoretical Nuclear Physics, National Laboratory
              of Heavy Ion Accelerator, Lanzhou 730000, China}
\author{Shan-Gui Zhou}%(ÖÜÉƹó)}%
 \email{sgzhou@itp.ac.cn}
%\homepage{http://www.itp.ac.cn/~sgzhou}
 \affiliation{State Key Laboratory of Theoretical Physics,
              Institute of Theoretical Physics, Chinese Academy of Sciences, Beijing 100190, China}
 \affiliation{Center of Theoretical Nuclear Physics, National Laboratory
              of Heavy Ion Accelerator, Lanzhou 730000, China}

\date{\today}

\begin{abstract}
The pseudospin symmetry is a relativistic dynamical symmetry 
connected with the small component of the Dirac spinor. 
The origin of pseudospin symmetry in single particle bound states in 
atomic nuclei has been revealed and studied extensively.
By examining the zeros of Jost functions corresponding to the small 
components of Dirac wave functions and phase shifts of continuum states,
we show that the pseudospin symmetry in single particle resonant states 
in nuclei is conserved when the attractive scalar and repulsive vector 
potentials have the same magnitude but opposite sign.
The exact conservation and the breaking of pseudospin symmetry are illustrated 
for single particle resonances in spherical square-well and Woods-Saxon potentials.
\end{abstract}

%03.65.Nk 	Scattering theory  
%03.65.Pm 	Relativistic wave equations  
%21.10.Hw 	Spin, parity, and isobaric spin  
%21.10.Pc 	Single-particle levels and strength functions  
%21.10.Tg 	Lifetimes, widths  
%21.60.Cs 	Shell model  
%21.60.Jz       Nuclear Density Functional Theory and extensions
%               (includes Hartree-Fock and random-phase approximations)
%24.10.Jv 	Relativistic models  
%25.70.-z 	Low and intermediate energy heavy-ion reactions  
%25.70.Ef 	Resonances  

\maketitle

%\end{CJK*}

The concept of pseudospin (PS) is often introduced to reveal the dynamical nature 
of quantum systems.
%e.g., quantum Hall ferromagnet and other quantum Hall systems, graphene, 
%Semiconductor Bilayers, pseudospin glasses. 
More than 40 years ago the pseudospin symmetry (PSS) in nuclear single 
particle states was observed:
PS doublets with quantum numbers $(n_r, l, j=l+1/2)$
and $(n_r-1, l+2, j=l+3/2)$ are nearly 
degenerate~\cite{Arima1969_PLB30-517, Hecht1969_NPA137-129}.
Since then much efforts had been devoted to explore the origin of
the PSS (see, e.g., ~\cite{Dudek1987_PRL59-1405, Bahri1992_PRL68-2133, 
Blokhin1995_PRL74-4149}) 
until it was shown that the PSS in nuclei is a relativistic
symmetry which is exactly conserved when the scalar and
vector potentials have the same size but opposite sign, i.e.,
$\Sigma(r) \equiv S(r) + V(r) = 0$~\cite{Ginocchio1997_PRL78-436}.
However, this condition is never met in finite nuclei, 
because in this limit there are no bound nuclei any more. 
Later it was found that the PSS is exact under a less strict condition,
$d\Sigma(r)/dr = 0$, and to what extent the PSS is conserved
is related to the competition between the centrifugal barrier and the PS
orbital potential~\cite{Meng1998_PRC58-R628, Meng1999_PRC59-154}.
This condition can also not be met in realistic nuclei,
therefore experimentally one always finds that the PSS is broken.
The above mentioned two conditions also result in the spin symmetry (SS) 
in anti-nucleon spectra~\cite{Ginocchio1999_PR315-231, Zhou2003_PRL91-262501} which
is much better developed than the PSS in nuclear 
single particle spectra~\cite{Zhou2003_PRL91-262501, He2006_EPJA28-265}.
The SS and PSS have been studied extensively within the relativistic framework
in which static mean fields dominate,
including the PSS in deformed nuclei~\cite{Lalazissis1998_PRC58-R45, Ginocchio2004_PRC69-034303}
and the SS for $\bar\Lambda$ spectra in hyper-nuclei~\cite{Song2009_CPL26-122102}.
The relevance of the PSS in nucleon-nucleus and nucleon-nucleon scatterings has 
also been discussed~\cite{Ginocchio1999_PRL82-4599, Leeb2000_PRC62-024602,
Ginocchio2002_PRC65-054002, Leeb2004_PRC69-054608}.
The readers are referred to Ref.~\cite{Ginocchio2005_PR414-165} for a review
and Refs.~\cite{Typel2008_NPA806-156, Leviatan2009_PRL103-042502, 
Long2010_PRC81-031302R, Lisboa2010_PRC81-064324, 
Liang2011_PRC83-041301, Guo2012_PC85-021302R} for some recent progresses.

In recent years, there has been an increasing interest in the exploration of 
nuclear single-particle resonant states especially in the study of exotic nuclei with 
unusual $N/Z$ ratios~\cite{Yang2001_CPL18-196, Zhang2004_PRC70-034308, 
Zhang2008_PRC77-014312, Fedorov2009_FBS45-191, Zhou2009_JPB42-245001, 
Guo2010_PRC82-034318, Pei2011_PRC84-024311}. 
In these nuclei, the neutron (or proton) Fermi surface is close to the particle continuum, 
thus the contribution of the continuum and/or resonances is important~\cite{Vretenar2005_PR409-101, Meng2006_PPNP57-470}. 
The study of symmetries in resonant states is certainly an interesting topic.
There have been some investigations of the PSS in single particle 
resonances~\cite{Guo2005_PRC72-054319, Guo2006_PRC74-024320, 
Zhang2006_HEPNP30S2-97, Zhang2007_CPL24-1199}.
For example, the PSS for the resonant states in $^{208}$Pb is investigated 
by solving the Dirac equation with Woods-Saxon-like vector and scalar potentials 
in combination with an analytic continuation in the coupling-constant 
method~\cite{Kukulin1989, Yang2001_CPL18-196, Zhang2004_PRC70-034308}
and it was found that the diffuseness of the potentials plays an important 
role in the splitting of energy and width of resonant PS partners~\cite{Guo2005_PRC72-054319}.
However, in all these studies, the PSS in resonances was investigated numerically 
and a rigorous justification of the PSS in single particle resonant states, 
like that for bound states given in Ref.~\cite{Ginocchio1997_PRL78-436}, 
is still absent.
In this Letter, we show that the PSS in single particle resonant states 
in nuclei is also exactly conserved under the same condition for the PSS in bound states, 
i.e., $\Sigma(r) = 0$ or $d\Sigma(r)/dr = 0$.  
We will also illustrate the exact conservation and the breaking of PSS 
in single particle resonances in spherical square-well and Woods-Saxon
potentials.

In a relativistic description, nuclei are characterized by a
strong attractive scalar potential $S(\bm{r})$ and a strong repulsive
vector potential $V(\bm{r})$~\cite{Vretenar2005_PR409-101, Meng2006_PPNP57-470}. 
The Dirac equation for a nucleon reads
\begin{equation}
 \left[ \bm{\alpha} \cdot \bm{p}
      + \beta \left( M + S(\bm{r}) \right)
      + V(\bm{r})
 \right] \psi(\bm{r})
 = \epsilon \psi(\bm{r})
 ,
~\label{Eq:diraceq}
\end{equation}
where $\bm{\alpha}$ and $\beta$
are the Dirac matrices and $M$ is the nucleon mass.
For a spherical nucleus, the Dirac spinor 
\begin{equation}
 \psi (\bm{r}) = \frac{1}{r}
  \left(
   \begin{array}{c}
    i F_{n\kappa}(r)         Y_{jm}^{l}(\theta,\phi )        \\
    - G_{\tilde{n}\kappa}(r) Y_{jm}^{\tilde{l}}(\theta,\phi )
   \end{array}
  \right) ,
 \label{eq:SRHspinor}
\end{equation}
where $Y_{jm}^{l}(\theta,\phi)$ is the spin spherical harmonic. 
$F_{n\kappa }(r)/r$ and $G_{\tilde{n}\kappa}(r)/r$ are the radial 
wave functions for the upper and lower components with $n$ and $\tilde{n}$ radial nodes.
$\kappa = (-1)^{j+l+1/2}(j+1/2)$
and $\tilde{l}=l-{\text{sign}}(\kappa)$.
The radial Dirac equation is then derived as
\begin{equation}
 \left(\begin{array}{cc}
                 M + \Sigma(r) & -\dfrac{d}{dr} + \dfrac{\kappa}{r}\\
         \dfrac{d}{dr}+\dfrac{\kappa}{r} & -M + \Delta(r)
       \end{array}
 \right)
 \left(\begin{array}{c}
         F(r)\\
         G(r)
       \end{array}
 \right)
 =  \epsilon  
 \left(\begin{array}{c}
         F(r)\\
         G(r)
       \end{array}
 \right)
 ,
~\label{Eq:radialdirac}
\end{equation}
where $\Sigma(r)$ is defined earlier, $\Delta(r) \equiv V(r)-S(r)$,
and $\epsilon$ is the eigenenergy.
For brevity we omit the subscripts from $F(r)$ and $G(r)$ whenever no confusion arises.
This first order coupled equation can be rewritten as two decoupled second order
differential ones. 
Here we only write down the one for the small component to which
the PSS is directly connected,
%\begin{widetext}
\begin{eqnarray}
 \left[
   \frac{d^{2}}{dr^{2}}
  -\frac{1}{M_{-}(r)}\dfrac{d\Sigma(r)}{dr}\frac{d}{dr}
  -\dfrac{\tilde{l}(\tilde{l}+1)}{r^{2}}
  \hspace*{1cm}
 \right.
 &   &
 \nonumber \\
 \left. \mbox{}
  + \dfrac{1}{M_{-}(r)}
    \frac{\kappa}{r}\dfrac{d\Sigma(r)}{dr}
  -  M_+(r) M_-(r)
 \right] G(r)
 & = & 
 0 , 
 \label{eq:G}
\end{eqnarray}
%\end{widetext}
where $M_{+}(r) \equiv M + \epsilon - \Delta(r)$ and $M_{-}(r) \equiv M - \epsilon + \Sigma(r)$.
Equation~(\ref{eq:G}) is fully equivalent to Eq.~(\ref{Eq:radialdirac}).

For the continuum in the Fermi sea, i.e., $\epsilon \geq M$, there exist two 
independent solutions for Eq.~(\ref{eq:G}). 
(Note that the following discussions are also valid for the continuum in the Dirac sea.)
The physically acceptable solution is the one that vanishes at the origin.
As usual we define the regular solution $G(r)$ as the one that behaves like 
$j_{\tilde{l}}(pr)$ as $r \rightarrow 0$~\cite{Taylor1972},
\begin{equation}
  \lim_{r \rightarrow 0} G(r)/j_{\tilde{l}}(pr) = 1,
  \ p = \sqrt{\epsilon^2 - M^2} .
 \label{eq:regularsolution}
\end{equation}

We now turn to the asymptotic behavior of the regular solution as $r\rightarrow\infty$. 
At large $r$ the potentials for neutrons vanish and the wave functions oscillate, 
Eq.~(\ref{eq:G}) becomes a Ricatti-Bessel equation with angular momentum $\tilde{l}$ and
the solution can be written as a combination of the
Ricatti-Hankel functions,
\begin{equation}
 G(r) 
 = \frac{i}{2} \left[
                       \mathcal{J}_{\kappa}^{G}(p)     h_{\tilde{l}}^{-}(pr) 
                     - \mathcal{J}_{\kappa}^{G}(p)^{*} h_{\tilde{l}}^{+}(pr)
               \right],
   \ r\rightarrow\infty
 ,
\end{equation}
where $\mathcal{J}_{\kappa}^{G}(p)$ is the Jost function for the
small component and $h_{\tilde{l}}^{\pm}(pr)$  the Ricatti-Hankel functions.

\begin{figure}
\begin{centering}
\includegraphics[width=0.9\columnwidth]{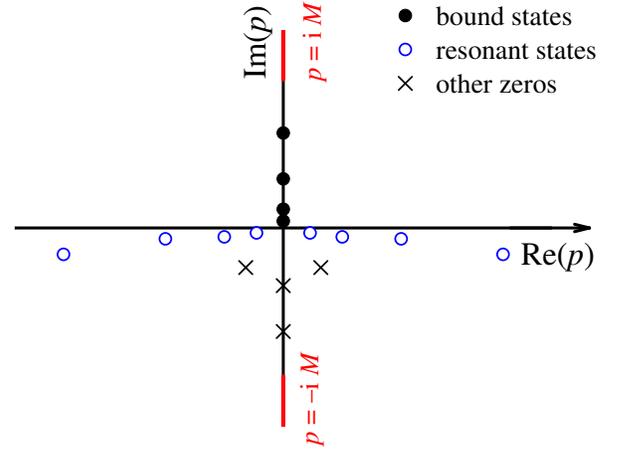}
\par\end{centering}
\caption{(Color online)
Schematic picture of the zeros of the Jost function $\mathcal{J}_{\kappa}^{G}$ on the 
complex momentum plane.
A cut is made on the imaginary axis, from $p=iM$ to infinity and back to $p=-iM$.
}
~\label{fig:zeros_sche}
\end{figure}

In nuclei the potentials $V(r)$ and $S(r)$ share some general properties,
e.g., they are analytic functions of $r$, vanish when
$r\rightarrow\infty$, and have no singularities. 
Under such conditions, the Jost function is an analytic function of $p$ 
and can be analytically continued to a large area in the complex $p$ plane. 
Here the structure of the $p$ Riemann surface on which the Jost functions are defined 
is more complex than the non-relativistic case. 
For example, the square root in the relativistic energy-momentum relation 
$\epsilon^2=p^{2}+M^{2}$ creates branching points at $p=\pm iM$, thus the 
corresponding Riemann surface is at least two folds.

In Fig.~\ref{fig:zeros_sche} the zeros of the Jost function 
$\mathcal{J}_{\kappa}^{G}(p)$ on the complex momentum plane
are schematically shown.
For simplicity we show only the first sheet with Re($\epsilon$)$\geq$0 which contains positive 
energy bound states and resonant states, while the other sheet with 
Re($\epsilon$)$\leq$0 can be used to investigate negative energy ones.
These two sheets are connected by a cut on the imaginary axis, 
from $p=iM$ to infinity and back to $p=-iM$.
Restricted to the first sheet and not too large $|p|$, $\mathcal{J}_{\kappa}^{G}(p)$ is a 
single valued analytic function of $p$.
The zeros of $\mathcal{J}_{\kappa}^{G}(p)$ are denoted by full circles (bound states), 
open circles (resonant states), and crosses (other zeros), respectively.
The zeros on the positive imaginary axis of the $p$ plane represent 
bound states of the original eigenvalue problem, 
while the zeros on the lower $p$ plane and 
near the real axis correspond to resonant states. 
The resonance energy $E_{{\rm res}}$ and width $\Gamma_{{\rm res}}$
are determined by the relation $E=E_{{\rm res}}-i{\Gamma_{{\rm res}}}/{2}=\sqrt{p^{2}+M^{2}}$. 
By examining the zeros of the Jost function we can study the bound
and resonant states on the same footing and 
many known properties of bound states can be generalized
to resonances straightforwardly.

In the PSS limit, Eq.~(\ref{eq:G}) is reduced as
\begin{equation}
 \left[
          \frac{d^{2}}{dr^{2}}
        - \frac{\tilde{l}(\tilde{l}+1)}{r^{2}}
        + \left(
                \epsilon-M
          \right)
          M_+(r)
%          \left(
%                \epsilon-\Delta(r)+M
%          \right)
 \right]  G(r)
 =  0
 .
~\label{eq:PSLeqG}
\end{equation}
For bound states it is an eigenequation that determines the eigenenergy $\epsilon$.
While for the continuum $\epsilon$ can be any value $\ge M$ and 
we mainly focus on wave functions and their asymptotic behavior.
For PS doublets with different quantum numbers $\kappa$ and $\kappa'$ with 
$\kappa^{\prime}=-\kappa+1$, the small components satisfy the same equation because 
they have the same pseudo-orbital angular momentum $\tilde{l}$~\cite{Ginocchio1997_PRL78-436}.
In particular, for continuum states, we have $G_{\kappa}(\epsilon,r)=
G_{\kappa^{\prime}}(\epsilon,r)$ for any energy $\epsilon$. 
Because the definition of the Jost function $\mathcal{J}_{\kappa}^{G}(p)$ only
depends on the asymptotic behavior of the small component, we have
$\mathcal{J}_{\kappa^{\prime}}^{G}(p)=\mathcal{J}_{\kappa}^{G}(p)$
on the positive real axis. 
This equivalence can be generalized into the complex $p$ plane 
due to the uniqueness of the analytic continuation. 
Thus the zeros are the same for $\mathcal{J}_{\kappa^{\prime}}^{G}(p)$ and 
$\mathcal{J}_{\kappa}^{G}(p)$: If there exists a resonant state with 
energy $E_{{\rm res}}$ and width $\Gamma_{{\rm res}}$ and the quantum number $\kappa$, 
there must be another one with the same energy and width and quantum number $\kappa^{\prime}$. 
That is to say, the PSS in single particle resonant states 
in nuclei is exactly conserved when the attractive scalar and repulsive vector 
potentials have the same magnitude but opposite sign.
Certainly if we focus on the zeros of the Jost functions of PS doublets
on the positive imaginary axis of the $p$ plane, 
we come to the well-known PSS for bound states.

In scattering theories, one can also determine resonance parameters
from the change of cross section or phase shift which 
give us more insights into the resonant phenomena. 
Next we discuss the PSS in resonant states by examining the phase shift. 
Using the asymptotic behavior of the Ricatti-Bessel functions, 
one obtains from Eq.~(\ref{eq:PSLeqG}),
\begin{equation}
            G_{\kappa}(r)  
  \propto   \sin  \left(
                         pr 
                       - \frac{\tilde{l}\pi}{2} 
                       + \delta_{\kappa}^{G}(p)
                  \right)
 ,\
 r\rightarrow\infty
 ,
 \label{eq:asymp}
\end{equation}
where the phase shift $\delta_{\kappa}^{G}(p)$ is related to the Jost function 
through $\mathcal{J}_{\kappa}^{G}(p)=|\mathcal{J}_{\kappa}^{G}(p)|e^{-i\delta_{\kappa}^{G}(p)}$.
Whenever $\delta_{\kappa}^{G}(p)=n\pi+\frac{\pi}{2}$, there is a resonant state 
and its width is determined by the tangent of the phase shift function $\delta_{\kappa}^{G}(p)$. 
In the PSS limit, the coincidence between $G_{\kappa}(r)$ and 
$G_{\kappa^{\prime}}(r)$ means that 
$\delta_{\kappa}^{G}(p)=\delta_{\kappa^{\prime}}^{G}(p)$ for any value of $p$. 
Therefore resonance parameters of PS doublets are the same.

It has been postulated in previous numerical studies that widths
of PS doublets should be different even in the PSS limit because 
centrifugal barriers are different for these two resonant states.
Here we have shown that this is not the case: 
For PS doublets of single particle resonant states in nuclei, not only
the energies, but also the widths are exactly the same in the PSS limit.

Similar to what happens in bound states, when the PSS limit is 
not realized, the PSS in resonant states is broken.
In the following we will use a solvable model
to illustrate the conservation and the breaking of the PSS in resonant states. 
Note that this kind of investigation can also be done numerically
with other potentials.

We consider that $\Sigma(r)$ and $\Delta(r)$
are both spherical square-well potentials,
\begin{eqnarray}
 \Sigma(r) & = & \left\{ \begin{array}{c}
                                C,\qquad r<R ,\\
                                0,\qquad r\geq R ,\end{array}
                 \right.  
 \label{eq:square_well_S}
 \\
 \Delta(r) & = & \left\{ \begin{array}{c}
                                D,\qquad r<R ,\\
                                0,\qquad r\geq R ,\end{array}
                 \right.
 \label{eq:square_well_D}
\end{eqnarray}
where $C$ and $D$ are constants and $R$ is the width.
For such potentials the wave function is continuous at $r=R$, 
but not its derivative
because the derivative of a square-well potential is a delta-function. 
For $r\neq R$, Eq.~(\ref{eq:G}) reads,
%\begin{widetext}
\begin{equation}
 \left[
           \frac{d^{2}}{dr^{2}}
        -  \frac{\tilde{l}(\tilde{l}+1)}{r^{2}}
        - M_+(r) M_-(r)
%                 +  \left(
%                           \epsilon - \Sigma(r) - M
%                    \right)
%                    \left(
%                           \epsilon - \Delta(r) + M
%                    \right)
          \right] G(r)
   =      0
 .
\end{equation}
%\end{widetext}
The regular solution of this equation is just a combination of the Ricatti-Bessel functions,
\begin{eqnarray}
 G(r) & = & (p/k)^{\tilde{l}+1} j_{\tilde{l}} \left( kr \right),\ r<R,
~\label{Eq:Gr-} 
 \\
 G(r) & = &   \frac {i}{2} \left[
                 \mathcal{J}_{\kappa}^{G}(p)     h_{\tilde{l}}^{-}(pr)
               - \mathcal{J}_{\kappa}^{G}(p)^{*} h_{\tilde{l}}^{+}(pr)
                            \right],
 r\geq R, %\nonumber \\
~\label{Eq:Gr+} 
\end{eqnarray}
with $k = \sqrt{\left( \epsilon - C - M \right) \left( \epsilon - D + M \right)}$.
The coefficient $(p/k)^{l+1}$ is inserted in accordance with Eq.~(\ref{eq:regularsolution})
and $\hat{j}_l(z) \propto z^{l+1}$ as $z \rightarrow 0$.

Next let us determine the Jost function from continuous conditions of
radial wave functions at $r=R$,
\begin{equation}
 G(R_{+}) = G(R_{-}), \quad  F(R_{+})=F(R_{-}) , 
~\label{Eq:connection}
\end{equation}
where $R_-$ and $R_+$ mean that one approaches $r=R$ from $r<R$ and $r>R$ respectively.
A linear equation for the Jost function can be written immediately 
using the connection condition for $G(r)$,
\begin{equation}
      \left( \frac{p}{k} \right)^{\tilde{l}+1} j_{\tilde{l}} \left( kR \right) 
  =   \frac {i}{2} \left[
                 \mathcal{J}_{\kappa}^{G}(p)     h_{\tilde{l}}^{-}(pR)
               - \mathcal{J}_{\kappa}^{G}(p)^{*} h_{\tilde{l}}^{+}(pR)
                            \right] 
 .
\end{equation}
The continuous condition for $F(r)$ can also be used to deduce a similar equation.
The derivative of $G(r)$ is not continuous at $r=R$,
\begin{equation}
  \left. \frac{dG}{dr} \right|_{R_+} 
- \left. \frac{dG}{dr} \right|_{R_-}
 =  - C F(R)
 .
~\label{Eq:GR_der}
\end{equation}
Approaching $R$ from $r<R$, we can represent $F(R)$ by
\begin{eqnarray}
 F(R) 
 & = & 
 \left.
 \frac{1}{M+C-\epsilon}
      \left( \frac{d}{dr} - \frac{\kappa}{r} \right) G(r)
 \right|_{r=R} 
 \nonumber \\
 & = & 
 \frac{(p/k)^{\tilde{l}+1}}{M+C-\epsilon}
 \left(
         kj_{\tilde{l}}^{\prime}(kR)
       - \frac{\kappa}{R} j_{\tilde{l}}(kR)
 \right)
 .
~\label{Eq:FatR}
\end{eqnarray}
By substituting Eqs.~(\ref{Eq:Gr-}), (\ref{Eq:Gr+}), and (\ref{Eq:FatR}) into 
Eq.~(\ref{Eq:GR_der}) we get a linear equation for 
$\mathcal{J}_{\kappa}^{G}(p)$ and $\mathcal{J}_{\kappa}^{G}(p)^{*}$.
Finally the Jost function for the small component reads,
%\begin{widetext}
\begin{eqnarray}
 \mathcal{J}_{\kappa}^{G}(p)  
 & = &  
 -\frac{p^{\tilde{l}}}{2ik^{\tilde{l}+1}}
  \left[
        j_{\tilde{l}}(kR) p h_{\tilde{l}}^{+\prime}(pR)
          -  kj_{\tilde{l}}^{\prime}(kR) h_{\tilde{l}}^{+}(pR)
  \right.
 \nonumber \\
 &   & \mbox{\hspace*{0.0cm}}
  \left.
          -  \frac{C}{\epsilon-M-C} \left(
                                           kj_{\tilde{l}}^{\prime}(kR)
                                         - \frac{\kappa}{R} j_{\tilde{l}}(kR)
                                    \right) h_{\tilde{l}}^{+}(pR)
  \right]
 .
 \nonumber \\
~\label{Eq:JostG}
\end{eqnarray}
%\end{widetext}
Now comparing Jost functions $\mathcal{J}_{\kappa}^{G}(p)$ and
$\mathcal{J}_{\kappa^{\prime}}^{G}(p)$, it is clear that they differ
only in the part containing $C$ because they have the same $\tilde{l}$.
In other words, in the PSS limit, i.e., $C=0$, we have 
$\mathcal{J}_{\kappa}^{G}(p)=\mathcal{J}_{\kappa^{\prime}}^{G}(p)$.
Consequently the PSS is conserved both in bound states and in resonant states.
If $C\ne 0$, the PSS is broken and we can study the PS splitting 
of the energy and the width for resonant states.
Due to the special form of the spherical square-well potentials,
the PSS-breaking term is separated from the PSS-conserving term in the Jost function, 
which makes the study of the conservation or the breaking of the PSS very convenient.

\begin{figure}
\begin{centering}
\includegraphics[width=1.0\columnwidth]{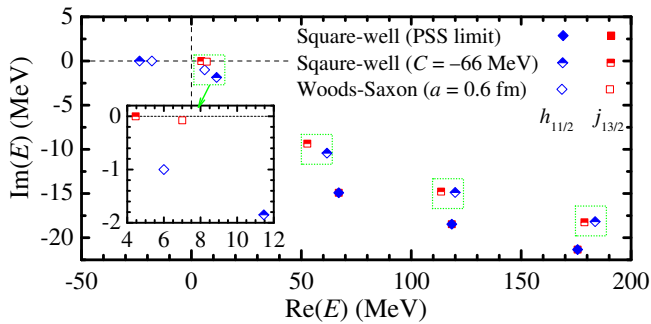}
\par\end{centering}
\caption{(Color online)
The zeros of the Jost function $\mathcal{J}_{\kappa}^{G}$ on the complex energy plane
in square-well potentials (\ref{eq:square_well_S}) and (\ref{eq:square_well_D}) with 
$C=0$ (solid symbols) and $C=-66$ MeV (half-filled symbols) 
for PS partners $h_{11/2}$ (diamond) and $j_{13/2}$ (square). 
The results with Woods-Saxon-like scalar and vector potentials are also
shown as open symbols.
}
~\label{fig:zeros}
\end{figure}

The solution of $\mathcal{J}_{\kappa}^{G}(p)=0$ can not
be written in a closed form. 
Here because the Jost function is analytic near its zeros, 
the secant method can be used for searching the roots. 
Starting from an initial guess for a root, the iteration converges after a few steps. 
In Fig.~\ref{fig:zeros} we show solutions in the complex energy plane 
for PSS doublets with $\tilde{l}=6$, i.e., $h_{11/2}$ with $\kappa = -6$ 
and $j_{13/2}$ with $\kappa' = 7$
for square-well potentials with $D=650$ MeV and $R=$ 7 fm.
In the PSS limit, i.e., $C=0$, all the roots locate in the lower 
half plane and there are no bound states. 
We show in Fig.~\ref{fig:zeros} three pairs of PS resonant partners by
full diamonds and squares. 
The conservation of the PSS for single particle resonant states is clearly seen.
When $C=-66$ MeV, there is one bound state only for $h_{11/2}$. 
Three pairs of PS partners of resonant states are shown by
half-filled diamonds and squares.
One finds the breaking of the PSS both in the bound states and in the resonant states.
For PS doublets with other values of $\tilde{l}$, we observed similar 
behaviors concerning the exact conservation and the breaking of the PSS.
We also studied resonances in Woods-Saxon-like potentials, 
$W(r) = W_0 / (1+\exp[(r-R)/a])$ ($W=V$ or $S$)
with parameters connected with $^{208}$Pb given in Ref.~\cite{Guo2005_PRC72-054319}: 
the depths $V_0 - S_0$ = 650 MeV and $V_0 + S_0$ = $-66$ MeV, the diffusivity parameter 
$a=0.6$ fm, and $R=7$ fm. 
Resonance parameters are obtained with the real stabilization method~\cite{Zhang2008_PRC77-014312}. 
The results are shown as open diamonds and squares for $h_{11/2}$ and $j_{13/2}$, respectively.
It is found that splittings of energy and width both become smaller
compared with the results with the square-well potentials.
The reason is that the derivative of $\Sigma(r)$ is smaller due to 
a non-zero diffusivity parameter.

Now we briefly discuss about protons. Due to the repulsive
Coulomb interaction, $\Sigma(r)$ can not be zero, nor its first derivative.
Therefore the PSS limit can never be realized for protons. However, the Coulomb
potential is relatively small compared with $\Sigma(r)$
and the breaking of the PSS from the Coulomb interaction should be small.

In summary, we show that the PSS in single particle
bound and resonant states in nuclei can be investigated on the 
same footing within the relativistic framework
by examining the zeros of Jost functions corresponding to
small components of nucleon Dirac wave functions.
In the PSS limit, i.e.,
the attractive scalar and repulsive vector 
potentials have the same magnitude but opposite sign,
small components of PS doublets are exactly the same. 
Thus Jost functions describing the asymptotic behavior
of the radial wave functions are identical to each other. 
When analytically continued to complex momentum plane, 
the resonant states, showing themselves as zeros of 
the Jost functions, are always paired in the PSS limit,
which leads to the exact PSS. 
The conservation of the PSS in the PSS limit
is also justified by examining the phase shift of continuum states. 
When leaving the PSS limit, the PSS in resonant states is broken.
These conclusions are tested for single particle resonances 
in spherical square-well and Woods-Saxon potentials.

%\acknowledgements
Helpful discussions with Peter Ring are acknowledged.
This work has been supported by 
Major State Basic Research Development Program of China (Grant No. 2007CB815000), 
National Natural Science Foundation of China (Grant Nos. 10875157, 10979066, 11175252, and 11120101005),
Knowledge Innovation Project of Chinese Academy of Sciences (Grant
Nos. KJCX2-EW-N01 and KJCX2-YW-N32). 
The results described in this paper are obtained on the ScGrid of
Supercomputing Center, Computer Network Information Center of Chinese Academy
of Sciences.

%\bibliographystyle{apsrev4-1}
%\bibliography{/home/bnlu/Desktop/Works/Documents/Personal.Files/Papers}
%\bibliography{../../../information/refs/JabRef/sgzhou}

%merlin.mbs apsrev4-1.bst 2010-07-25 4.21a (PWD, AO, DPC) hacked
%Control: key (0)
%Control: author (8) initials jnrlst
%Control: editor formatted (1) identically to author
%Control: production of article title (-1) disabled
%Control: page (0) single
%Control: year (1) truncated
%Control: production of eprint (0) enabled
%

\end{document}